\documentclass[10pt]{article}
\usepackage{txfonts}
\usepackage{graphicx}
\usepackage{natbib}
\bibpunct{(}{)}{;}{a}{}{,} 

\newcommand{\EQ}{\begin{equation}}
\newcommand{\EN}{\end{equation}}
\newcommand{\EQA}{\begin{eqnarray}}
\newcommand{\ENA}{\end{eqnarray}}
\newcommand{\Fig}[1]{Fig.~\ref{#1}}
\newcommand{\Eq}[1]{Eq.~(\ref{#1})}
\newcommand{\Sec}[1]{Section~\ref{#1}}
\newcommand{\Tab}[1]{Table~\ref{#1}}
\newcommand{\BB}{{\bf {B}}}
\newcommand{\JJ}{{\bf {J}}}

\newcommand{\aap}{A\&A}

\newcommand{\apjl}{ApJ}

\begin{document}

\title{Current accumulation at an asymmetric 3D null point caused by generic shearing motions}

\author{K. Galsgaard\footnote{Niels Bohr Institute, Blegdamsvej 17, Dk-2100 Copenhagen {\O}, Denmark} \and D. I. Pontin\footnote{Division of Mathematics, University of Dundee, Dundee, DD1 4HN, United Kingdom}}
\date{\today}

\maketitle

\begin{abstract}{{\it Context.}
Here we investigate the dynamical evolution of the reconnection process at an initially linear 3D null point that is stressed by a localised shear motion across the spine axis. The difference to previous investigations is that the fan plane is not rotationally symmetric and this allows for different behaviours depending on the alignment of the fan plane relative to the imposed driver direction.}\\
{{\it Aims.} The aim is to show how the current accumulation and the associated reconnection process at the non-axisymmetric null depends on the relative orientation between the driver imposed stress across the spine axis of the null and the main eigenvector direction in the fan plane.}\\
{{\it Methods.} The time evolution of the 3D null point is investigated solving the 3D non-ideal MHD equations numerically in a Cartesian box. The magnetic field is frozen to the boundaries and the boundary velocity is only non-zero where the imposed driving is for stressing the system is applied.}\\
{{\it Results.}The current accumulation is found to be along the direction of the fan eigenvector associated with the smallest eigenvalue until the direction of the driver is almost parallel to this eigenvector. When the driving velocity is parallel to the weak eigenvector and has an impulsive temporal profile the null only has a weak collapse forming only a weak current layer. However, when the null point is stressed continuously boundary effects dominates the current accumulation.}\\
{{\it Conclusions.} There is a clear relation between the orientation of the current concentration and the direction of the fan eigenvector corresponding to the small eigenvalue. This shows that the structure of the magnetic field is the most important in determining where current is going to accumulate when a single 3D null point is perturbed by a simple shear motion across the spine axis. As the angle between the driving direction and the strong eigenvector direction increases, the current that accumulates at the null becomes progressively weaker.}
\end{abstract}


\maketitle

\section{Introduction}
Magnetic reconnection is one of the main processes responsible for releasing free magnetic energy in complicated stressed magnetic field. A better knowledge of this process therefore helps us to better understand the possible non-ideal dynamical evolution of such magnetic fields. One such example is the magnetic field filling the photosphere - corona region. It has a highly complicated structure with variations on all possible length scales. It evolves dynamically, both through slow ideal changes that maintain the topology of the magnetic field and sometimes through fast, local, non-ideal processes where the stressed magnetic field releases a significant fraction of its free energy. Astrophysical magnetic fields are three-dimensional (3D) and therefore to understand the reconnection process, we have to investigate generic 3D magnetic fields. In 3D space magnetic reconnection can take place under two generally different circumstances \citep{1988JGR....93.5547S}, either in regions where the magnetic field vector is not vanishing, or at 3D null points and their associated separators. The first type of reconnection occurs in magnetic field regions associated with rapid, but continuous, changes in field line mapping. These regions typically allow for small field perturbations to generate strong current concentrations far away along the perturbed field lines, and are typically identified as being associated with Quasi-Separatrix Layers (QSLs) \citep{1995JGR...10023443P}. This has been disputed in more recent investigations \citep{2010A&A...516A...5W,2011A&A...525A..57P} where they show that QSLs and current accumulations do not necessarily represent the same magnetic field line region. Reconnection may also take place in magnetic fields with nulls, but not only at the null points. When no null reconnection takes place in these cases it is found to do so along magnetic separator lines connecting two null points \citep{1996JGR...10113445G,2010ApJ...725L.214P}. Finally, reconnection can also proceed directly at magnetic null points through a number of different typical evolution processes \citep{2009PhPl...16l2101P}. Such nulls have recently be shown to be present in abundance in the solar corona \citep{regnier2008,longcope2009}. Aspects of magnetic reconnection at a single rotationally symmetric 3D null point have been investigated by a number of authors \citep{2007PhPl...14e2109P, 2007JGRA..11203103P, 2011A&A...529A..20G, 2003JGRA..108.1042G, 2010SSRv..tmp...62M}. Here the spine axis is typically perturbed by external stress and the null reacts by generating a current concentration in its vicinity leading to a typical spine-fan reconnection scenario \citep{2009PhPl...16l2101P,2007PhPl...14e2106P}. To extend these investigations, \citet{2010A&A...512A..84A} looked at how current accumulates at a null that is not rotationally symmetric. They found the degree of asymmetry to have strong implications for the geometry of the current accumulation. As the asymmetry of the null increases the current sheet extends steadily further along the eigenvector of the numerically smallest (`weak') fan eigenvalue (hereafter referred to as the `WE' direction). In their investigation, the stress of the spine axis was directed along the dominating fan eigenvector direction, and therefore gives no indication to the collapse of the null and the formation  of the current sheet as the orientation between the dominating fan eigenvector direction (SE) and the imposed stress of the spine axis changes. In a different experiment, \citet{1997ASPC..111...82G} investigated the behaviour of a double null pair, showing how the presence of a separator connecting the two null points has a clear local focusing effect on the current independent of the relative orientation of the imposed driver and the separator. At the same time it was found that on the opposite side to the separator, the current accumulation changed to become perpendicular to the imposed driving direction. Despite the difference in the topological structures between a double null point pair and a single null, this strongly suggests that the asymmetric structure of a null will be able to significantly influence the alignment of the current accumulation. This is the topic investigated in this paper. A simple potential 3D null point is chosen, where a fixed ratio between the three eigenvalues (all different) is chosen. The influence of the relative orientation of the fan eigenvectors and the imposed driving on the dynamical response of the null point structure is investigated.

The layout of the paper is as follows. In \Sec{setup.sec} we discuss the structure of the magnetic field and the imposed driving profiles in relation to the numerical domain. Sections \ref{exp.sec} and \ref{impulsive.sec} describe the results of the different experiments, combining both impulsive driving and a long time consistent stressing of the system. \Sec{discu.sec} discusses the implications of the findings, and finally \Sec{conc.sec} summaries the conclusions of the investigation. 

\section{Numerical setup}
\label{setup.sec}
\begin{figure}
{\hfill \includegraphics[width=0.45\textwidth]{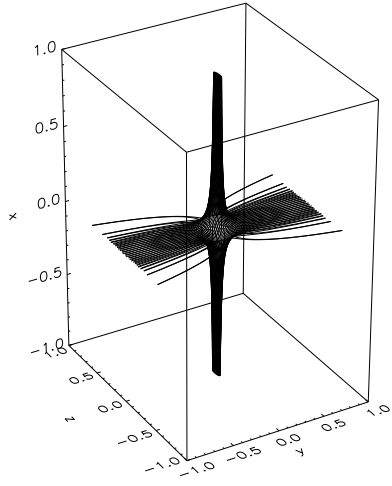} \hfill}
\caption[]{\label{init.fig} The structure of the initial potential 3D null point for $\theta=0$. The $x$-direction represents the spine axis, the $y,z$-axes the eigenvector direction in the fan plane. Here the $z$-direction represents the numerically smaller eigenvalue.}
\end{figure}

The magnetic field configuration adopted here is the linear rotationally asymmetric potential magnetic field defined by
\EQA
\BB &=& B_0 [-2x,\\
&&z\cos(\theta)\sin(\theta)(2-2\kappa) + y((2-\kappa)\cos(\theta)^2 + \kappa\sin(\theta)^2), \nonumber \\
&&z(\kappa \cos(\theta)^2 + (2-\kappa) \sin(\theta)^2) + y(2-2\kappa)\cos(\theta)\sin(\theta)], \nonumber
\ENA
where $B_0$ is a scaling parameter, here of order unity, $\kappa$ is a number between 0 and unity determining the asymmetry of the null point -- here we take $\kappa=0.5$ -- and $\theta$ is the angle between the y-axis and the major fan eigenvector. (The eigenvalues for the eigenvectors defining the fan plane of the null are ((2-$\kappa$), $\kappa$).). The null point is located at $(x,y,z)=(0,0,0)$ with the spine axis along the $x$-axis and the fan plane represented by the $y-z$ plane, i.e. $x=0$. The structure of the null point for $\theta=0$ is shown in \Fig{init.fig}. To investigate the dynamical evolution of the 3D null, the domain of interest is limited to $(x,y,z) = \pm (0.5,1.5,1.5)$, and has been discretised using a uniform grid with $200x300^2$ grid-points.

The plasma is initially assumed to be at rest and have a constant density and thermal energy. In the present experiment the MHD equations are non-dimensionalised and the density is set to unity, while the thermal energy is set to 0.025.

The linear nature of the magnetic null point forces a limitation on the numerical domain size. As a Cartesian domain does not represent very well the asymmetry of the magnetic null structure as it is rotated around the $x$-axis, and because the imposed boundary stressing also breaks any symmetry in the 3D null point, it is a non-trivial problem to describe the boundary conditions as the perturbation reaches the boundaries. In this approach, the flow perpendicular to all of the boundaries is imposed to be zero, and parallel velocities at the boundaries are restrict to be non-zero only in the two regions where dedicated stressing velocity flows are imposed. 

\begin{figure}
{\hfill \includegraphics[width=0.45\textwidth]{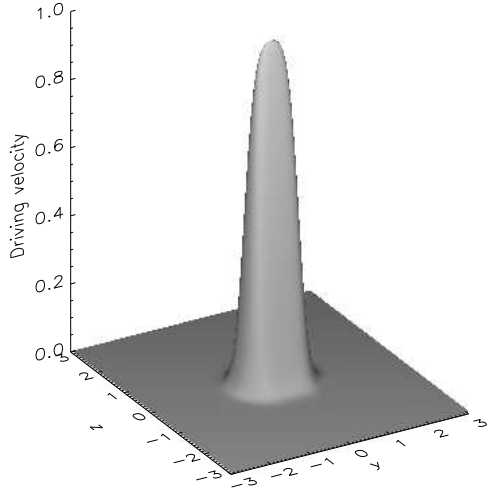} \hfill}
\caption[]{\label{driver.fig} A surface plot of the imposed normalised driving velocity profile.}
\end{figure}

The driving of the system is limited to two areas on the $x-$boundaries, by advecting the magnetic field in the $y$-direction, with an additional spatial dependence on the $z$-direction. The imposed driver has the following mathematical expression,
\EQA
\label{driver.eq}
V_y(y,z,t) & =  & \pm 0.5\,V_d(t)\,(\tanh((y-y_0)/y_h)-\tanh((y+y_0)/y_h)) \\
         & &             ~~~~~~ \times~ (\tanh((z-z_0)/z_h)-\tanh((z+z_0)/z_h)) , \nonumber
\ENA
where $V_d(t)$ is a time dependent profile related to the specific experiment discussed below, and $(y,z)=(\pm y_0,\pm z_0)$ represent the locations (at the centre of the hyperbolic tangent function) where the driving velocity has half of its peak value. The variables $y_h$, $z_h$ are the half widths of the hyperbolic tangent functions. This gives a near constant driving velocity between $(y,z)=(\pm(y_0-y_h), \pm(z_0-z_h))$ while the velocity decreases outside this region, asymptotically approaching zero. For these experiments we use $y_0 = 0.6$, $y_h = 0.2$, $z_0 = 0.3$ and $z_h = 0.2$. This profile is imposed on the two $x$-boundaries with opposite directions. Two types of driving are used. In the first the driving is switched on at the initiation of the experiment and reaches the peak velocity exponentially with a given short time scale. The driving is then maintained at this amplitude for the full extent of the experiment. For the second series of experiments, the same spatial driving profile is used, but this time is only imposed for a limited time to give instead an impulsive perturbation of the system. For both cases the structure of the driving profile relative to the domain extent in the $y-z$ directions is seen in \Fig{driver.fig}.

The experiments are conducted using the 3D non-ideal MHD code by \citet{Nordlund+Galsgaard97}. This is a high order finite difference code using staggered grids to maintain conservation of the physical quantities. The interpolation operators are fifth order in space while the derivative operators are sixth order. The solution is advanced in time using a third order explicit predictor-corrector method. A constant $\eta$ model for the resistivity is assumed, with a non-dimensional value of $10^{-3}$.

\section{Continued driving}
\label{exp.sec}

\begin{table}
\caption[]{\label{driver.tab} Experiments investigating the influence of the current accumulation and reconnection as the angle between the driver and the null point's symmetry axis is changed.
} 
\begin{tabular}{ c c c c c }
\hline
\hline
Name$^{~\mathrm{a}}$ & $\theta^{~\mathrm{b}}$ & Peak $J^{~\mathrm{c}}$ & Peak $V^{~\mathrm{d}}$ & $\Psi^{~\mathrm{e}}$ \\ 
\hline   
A  &  0 & 36.6 & 0.70 & 0.019 \\ 
B  & 30 & 34.5 & 0.62 & 0.017 \\ 
C  & 60 & 28.7 & 0.43 & \\
D  & 70 & 25.4 & 0.40 & \\
E  & 80 & 20.5 & 0.40 & \\ 
F  & 90 & 20.9 & 0.43 & \\ 
IA &  0 & 6.4 & 0.078 & 0.000855 \\
IB & 15 & 6.3 & 0.076 & 0.000755 \\
IC & 30 & 5.9 & 0.071 & 0.000695 \\
ID & 45 & 5.2 & 0.061 & 0.000615 \\
IE & 60 & 4.3 & 0.047 & 0.000560 \\
IF & 75 & 3.2 & 0.039 & 0.000465 \\
IG & 90 & 2.7 & 0.038 & 0.000390 \\
\hline
\end{tabular}
\begin{list}{}{}
\item[$^{\mathrm{a}}$] First five represent continued stressing (A-F) while the last seven (IA-IG) represents the impulsive driving.
\item[$^{\mathrm{b}}$] Angle between the weak fan eigenvalue and the z-axis.
\item[$^{\mathrm{c}}$] Peak current value.
\item[$^{\mathrm{d}}$] Peak outflow velocity.
\item[$^{\mathrm{e}}$] Peak integrated parallel electric field.
\end{list}
\end{table}

This section discusses the case of the continuous driving. For this case the driving amplitude has been modulated in the following way to;
\EQ
V_d=v_0 \tanh(t/t_{rise}), 
\EN
to smoothly ramp up the driving velocity to a constant amplitude. In these experiments a ramp up time of $t_{rise}=0.1$ has been used.
A number of experiments are conducted in which the angle between the direction of the driver and the orientation of the main fan eigenvector is changed between 0 to 90 degrees, experiments A to F. To quantify the main changes to the dynamical evolution of the magnetic field close to the null point a number of measurable numbers are chosen. These are listed in \Tab{driver.tab} and a discussion of each of them is provided below. 
To simplify our discussion we define here two orthogonal directions in the $yz$-plane (the fan plane). The first is the direction of the eigenvector associated with the strongest eigenvalue, denoted the `SE' direction, while we denote the direction of the eigenvector associated with the weak eigenvalue by `WE'. 

As the driver is initiated, a pulse propagates along the field lines towards the null region. Due to couplings between the magneto-acoustic waves the pulse does not propagate as a pure Alfv{\'e}n wave and it can therefore concentrate on the null point, eventually creating a local current concentration at the null. As the system is continuously stressed the current magnitude increases and eventually degenerates the null forming a localised current concentration, at least for the majority of the conducted experiments. 

\subsection{Current accumulation}

\begin{figure}
{\hfill \includegraphics[width=0.4\textwidth]{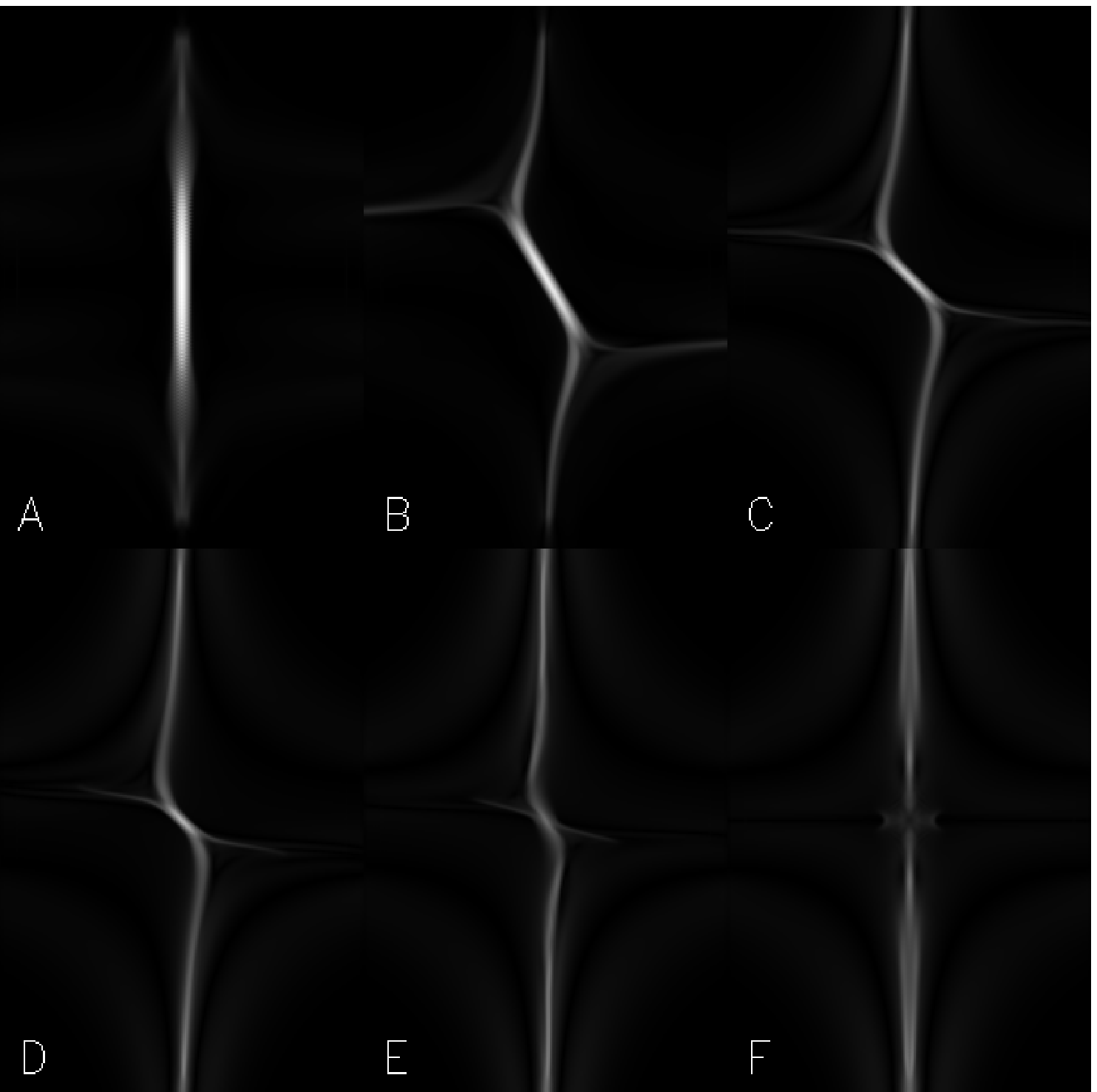} \hfill}

{\hfill \includegraphics[width=0.4\textwidth]{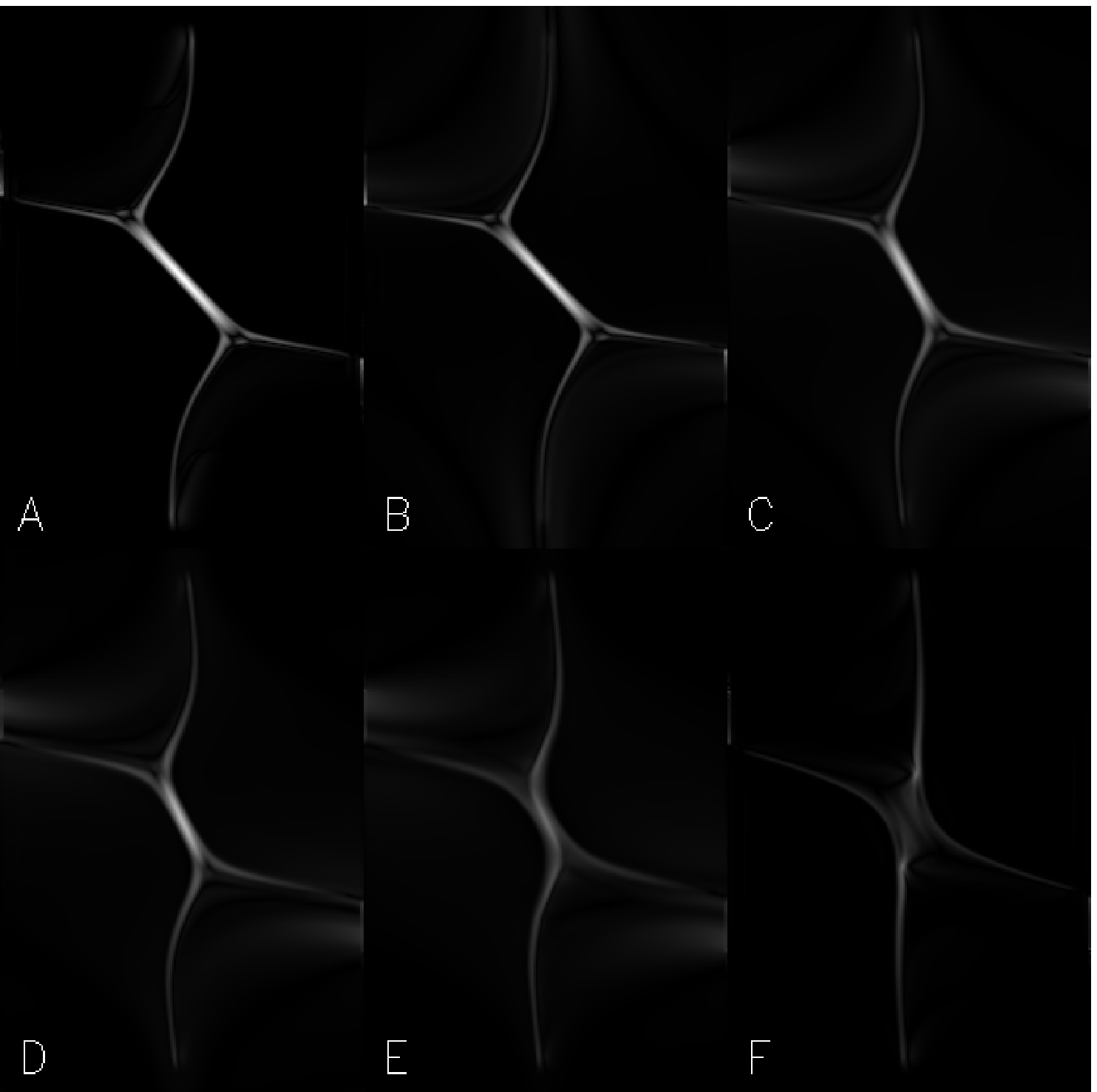} \hfill}

{\hfill \includegraphics[width=0.4\textwidth]{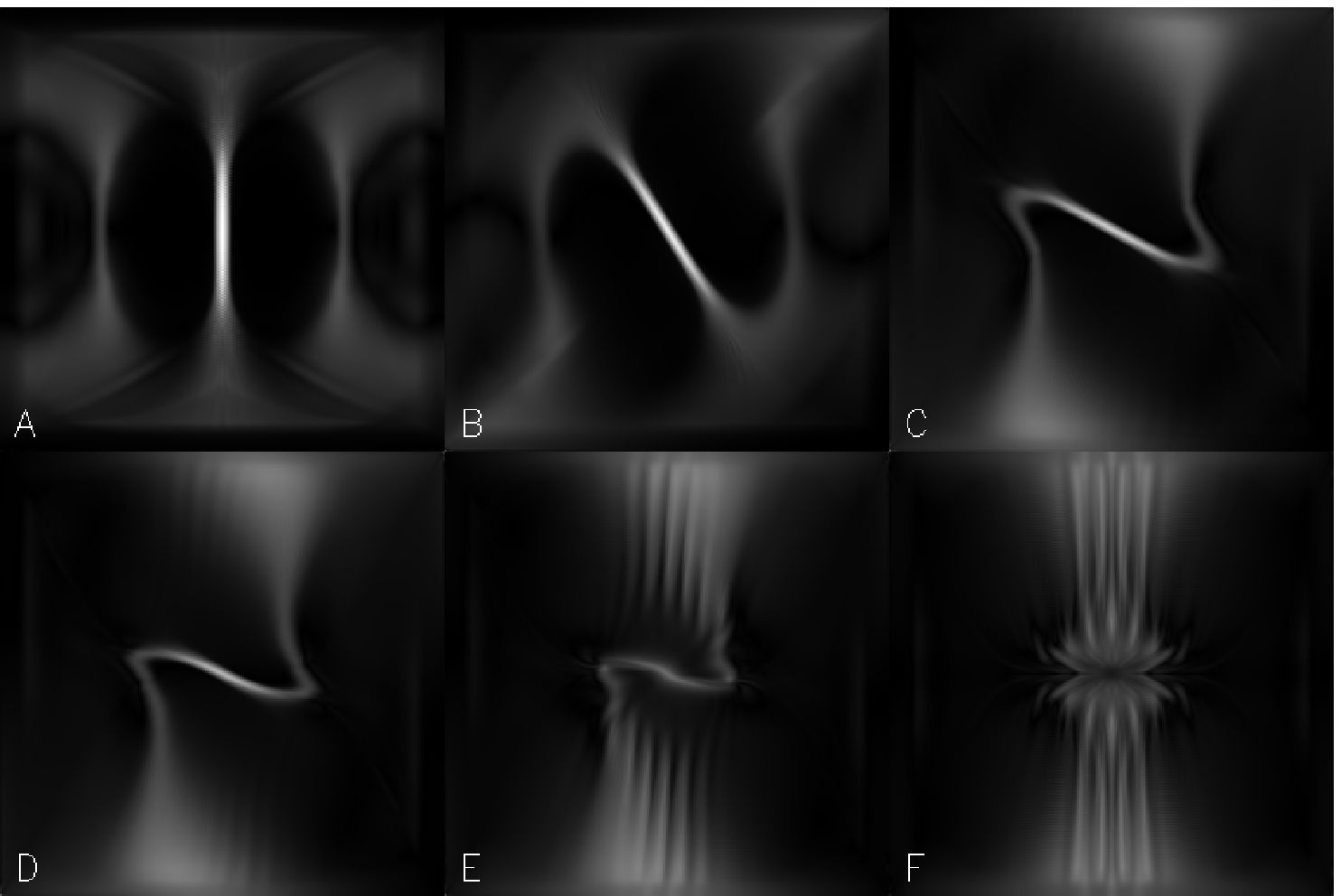} \hfill}

\caption[]{\label{current_structure.fig} Images of the current structure for experiments A-F in \Tab{driver.tab} at t=7.5. The top frame shows the $xy$-plane, the middle frame represents the $xz$-plane and finally the lower frame shows the $yz$-plane. In all frames the redundant coordinate is zero and the initial null point is therefore located at the centre of the images. The dynamical range in each of the frames are scaled according to the absolute peak current of the experiments.}
\end{figure}
\begin{figure}
{\hfill \includegraphics[width=0.5\textwidth]{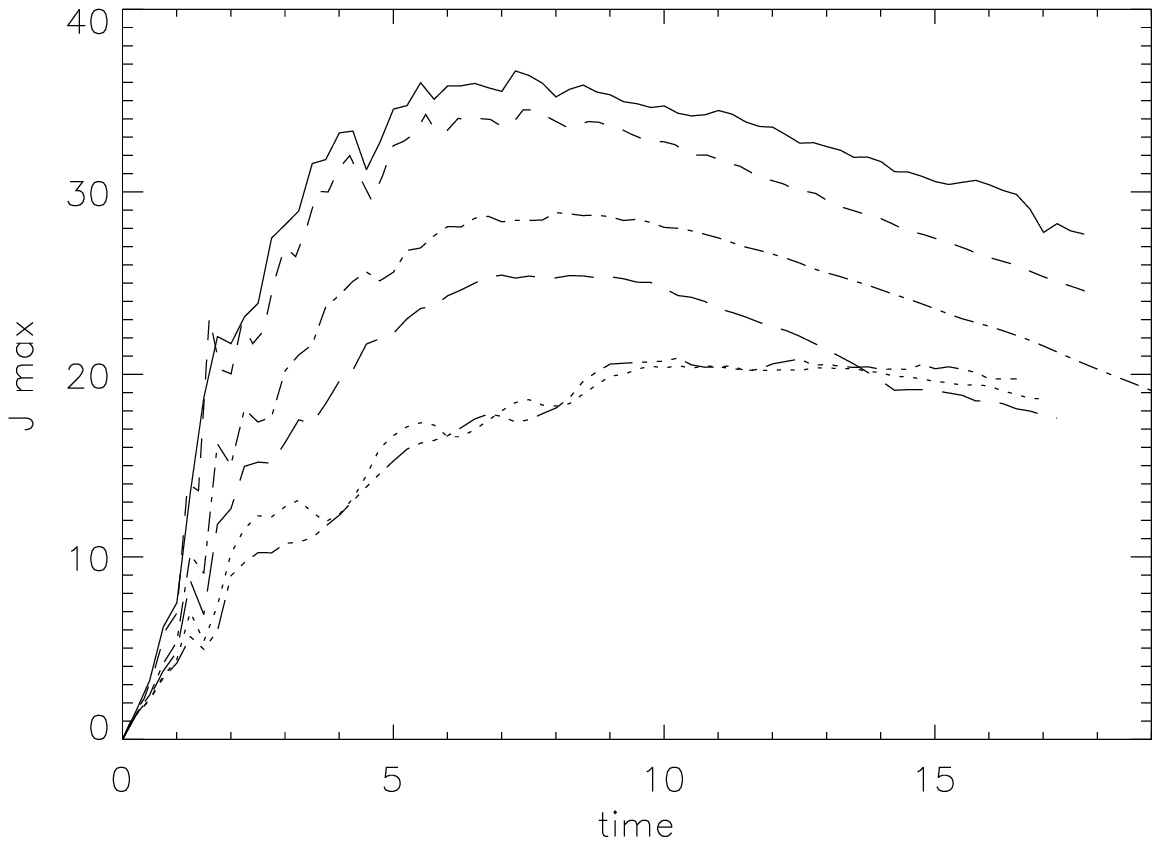} \hfill}
\caption[]{\label{current_evolution.fig} The time evolution of the peak current for the six continuously driven experiments. The lines represent experiment: A: full line, B: short dashed line, C: dot-dashed line, D: long dashed, E: dotted, F: triple dotted dashed line.}
\end{figure}

The frames in \Fig{current_structure.fig} show how the spatial orientation of the current sheet changes for different $\theta$ values. For $\theta=0$ the current sheet is clearly tilted in the $xy$-plane (the plane perpendicular to the shear driving), while extending a comparable distance in the $z$-direction, such that it approximates a tilted circular disk. As $\theta$ increases the orientation of the current sheet changes, always aligning itself along the WE direction. At the same time the width (in the SE direction) of the current sheet decreases making it appear increasing long. When $\theta$ is larger then 70 degrees, implying that the WE direction becomes almost parallel with the imposed driver, no significant current accumulates at the null point. Instead the strongest current is located close to the $z$-boundaries. This is most easily seen in the middle frame. The lower frame shows a transition in the current accumulation in the $yz$ plane. For $\theta=0$ the current connects in a straight line between the two $z$-boundaries, while for increasing $\theta$ it starts connecting with secondary current structures forming an $S$-like structure. This structure vanishes for experiment $E$ and $F$. The effect of the driving is therefore different for these experiments where the driver is closest to being aligned with the WE direction. We return to this in the discussion of the velocity flows below.

\Fig{current_evolution.fig} shows the time dependence of the peak current in the A-F experiments, measured in a small region around the initial null position. This shows that typically the peak current magnitude slowly builds, reaching its peak value around t=7.5. The peak current decreases with increasing angle, $\theta$, a similar behaviour is seen for the time where the different experiments reach their peak current for experiments $A-D$. Again experiments $E$ and $F$ behave differently, reaching the peak value later without decreasing again. This is because the current gradually extends inwards from its maxima at the $z$-boundaries, eventually entering the measurement region around the null. 

\subsection{Velocity flow}
\begin{figure}
{\hfill \includegraphics[width=0.5\textwidth]{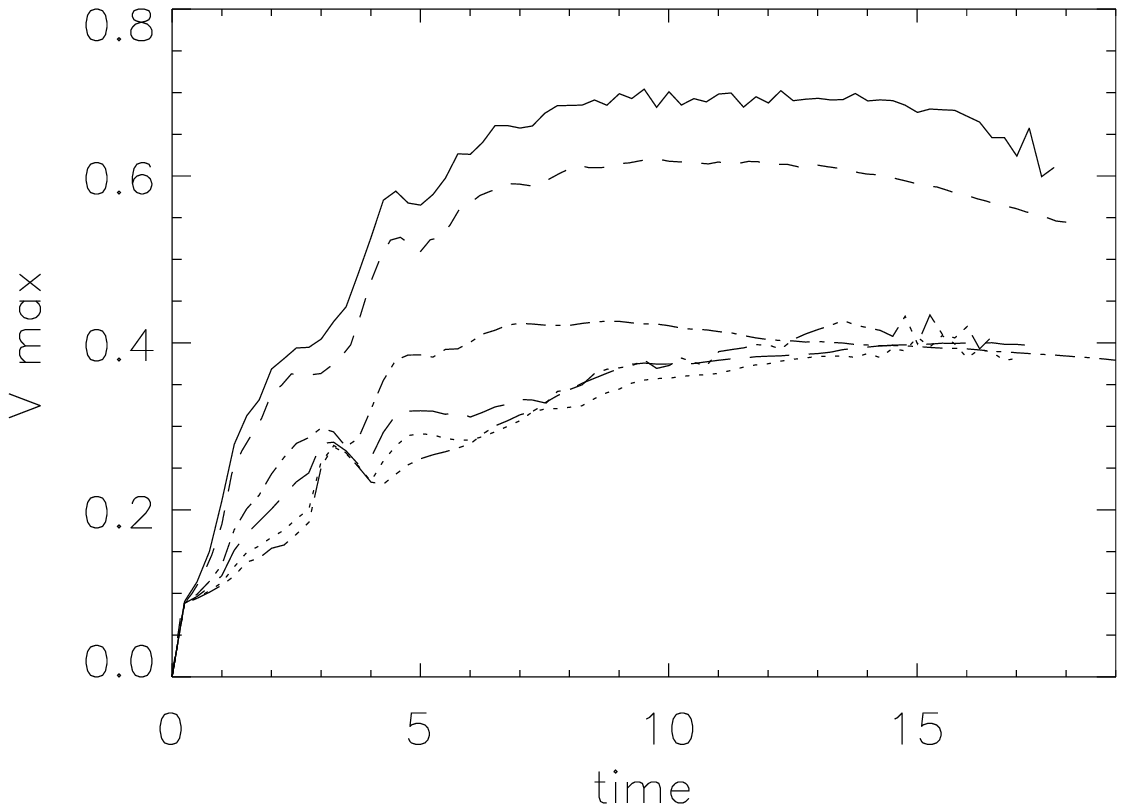} \hfill}
\caption[]{\label{velocity_evolution.fig} The time evolution of the peak jet velocity for the six experiments. The lines represents experiment: A: full line, B: short dashed line, C: dot-dashed line, D: long dashed, E: dotted, F: triple dotted dashed line.}
\end{figure}
\begin{figure}
{\hfill \includegraphics[width=0.35\textwidth]{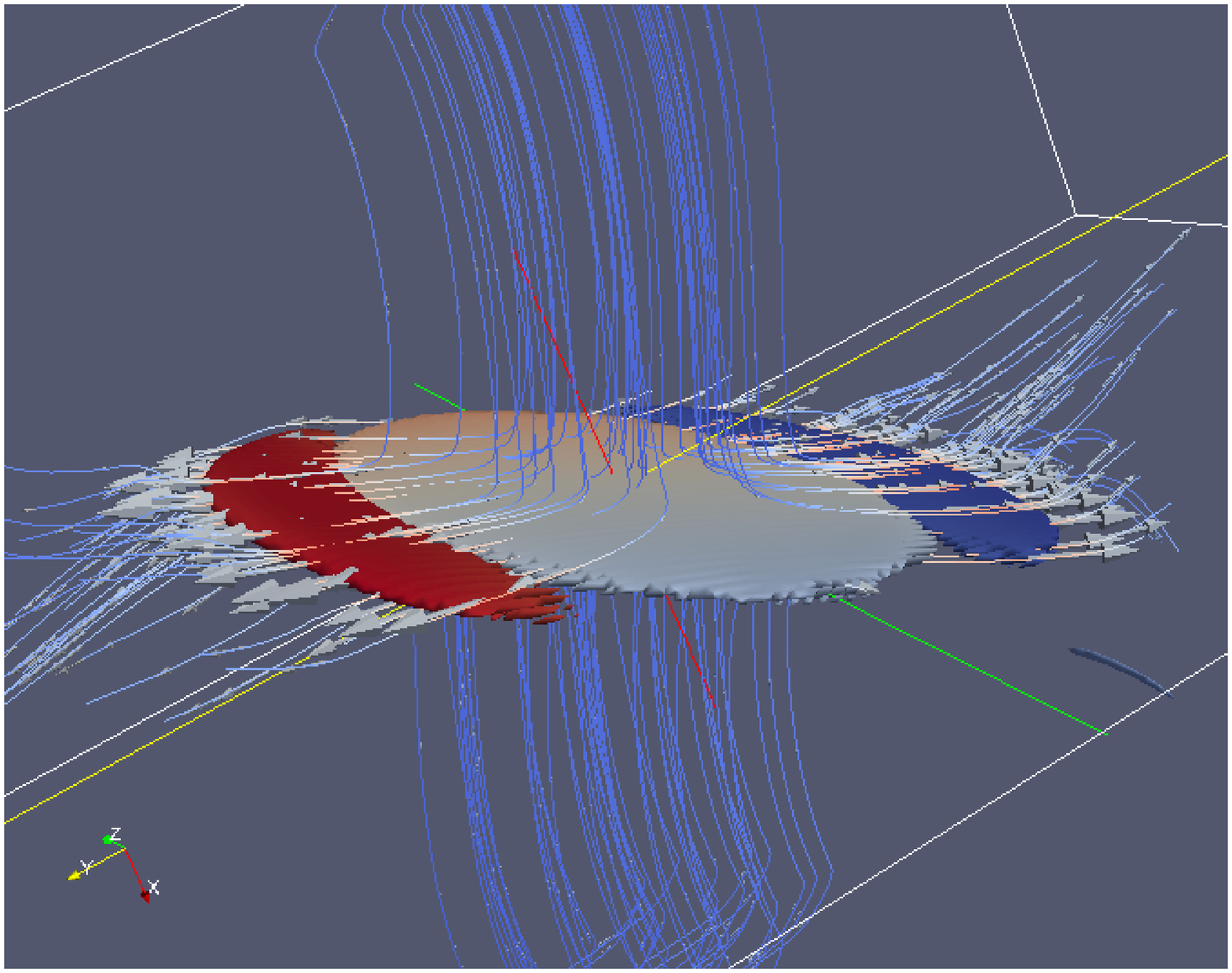} \hfill}

{\hfill \includegraphics[width=0.35\textwidth]{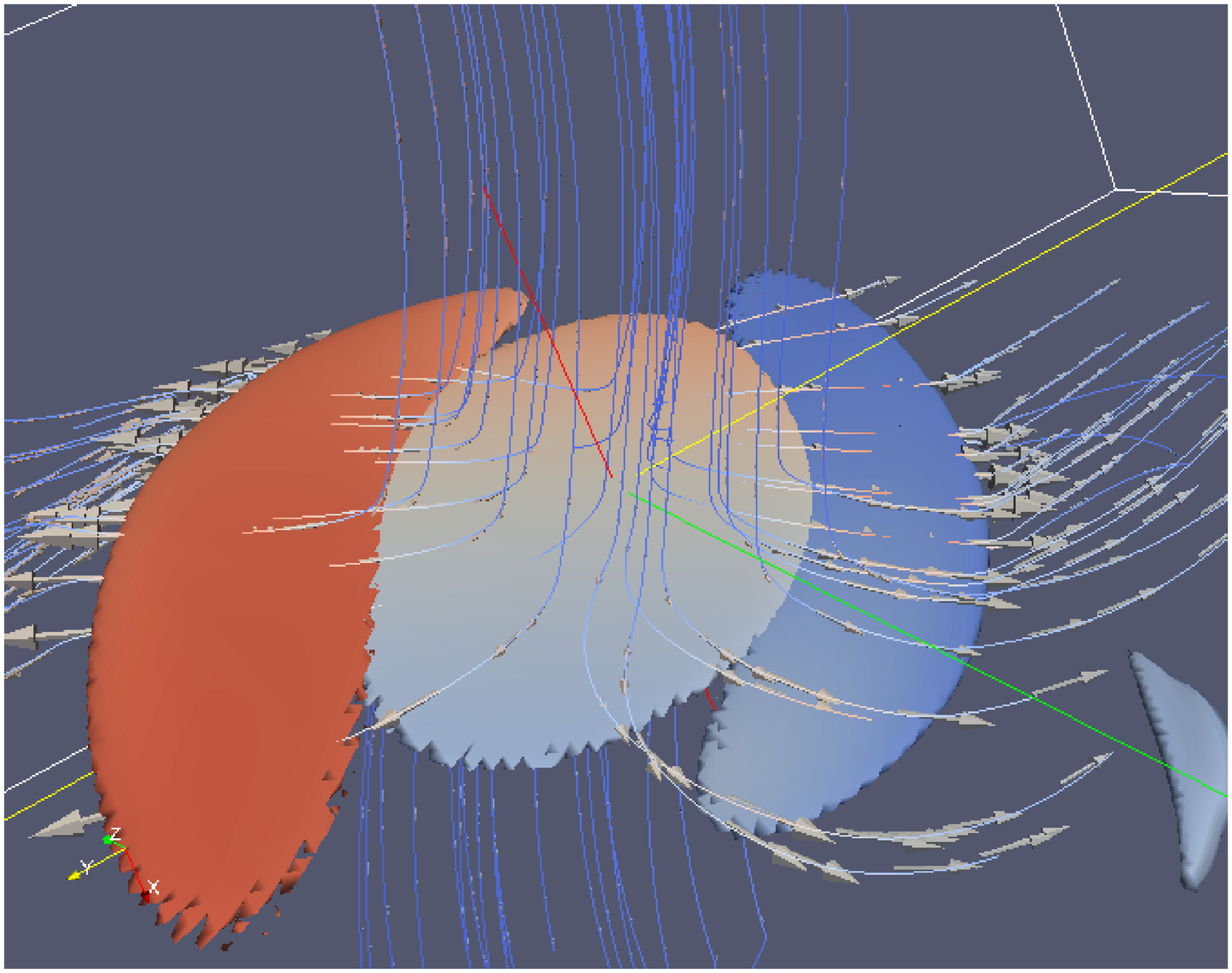} \hfill}

{\hfill \includegraphics[width=0.35\textwidth]{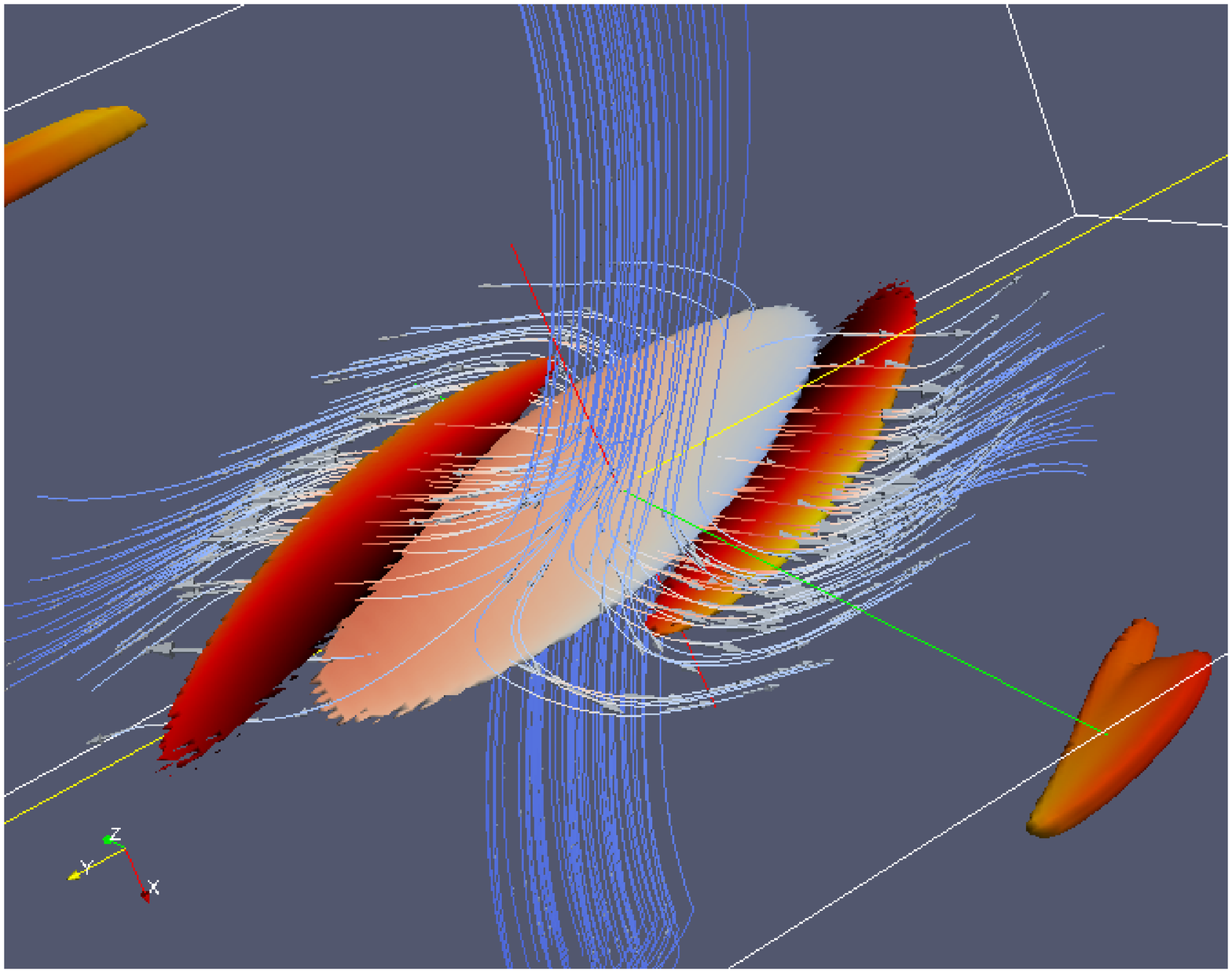} \hfill}

{\hfill \includegraphics[width=0.35\textwidth]{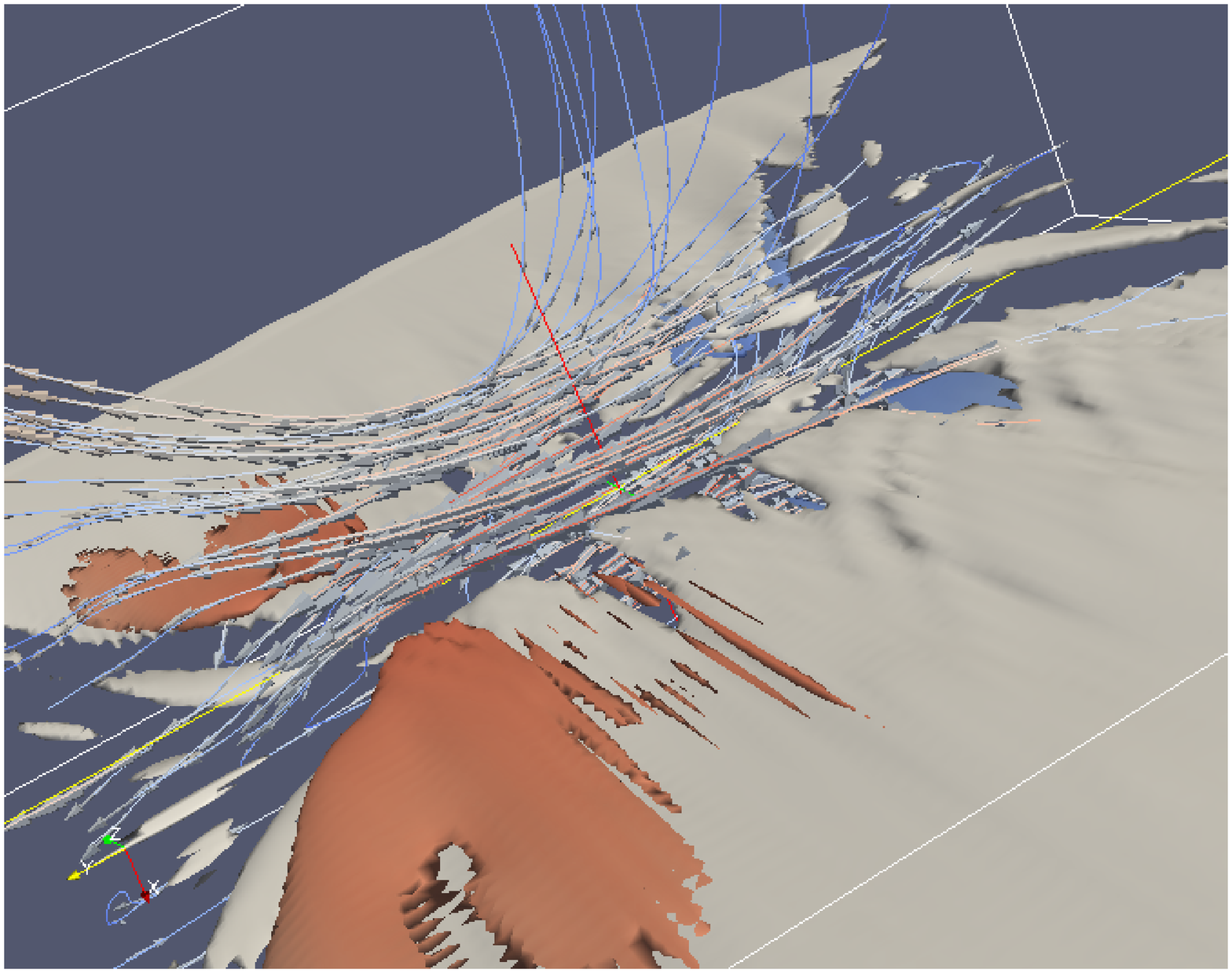} \hfill}

\caption[]{\label{velocity_struct.fig} The four frames show the structure of the reconnection region at t=7.5 for experiments A, B and C while F is shown at t=10.0. The field lines represent the velocity streamlines, with the arrows indicating the local strength of the velocity. The central isosurface represents the current sheet, while the two adjacent isosurfaces shows the spatial location with the highest flow velocities -- the reconnection jet.}
\end{figure}

\Fig{velocity_evolution.fig} shows the time evolution of the peak velocity of the plasma, measured in a small region around the initial null position. The velocity has a similar dependence on $\theta$ as the peak current discussed above, with the peak velocity decreasing with increasing $\theta$. The traditional delay between peak current and peak velocity is seen for all experiments, representing the time delay required for accelerating the plasma out of the diffusion region. The relative increase between experiments is larger for the velocity than for the current, indicating that the effective combination of the Lorentz force and the gas pressure decreases with increasing angle. Looking at the graph, it is seen that the peak velocities in experiments $D$, $E$ and $F$ follow nearly the same evolution, reaching a peak value later then the three other experiments. This is in contrast to the peak current, where only $E$ and $F$ had similar evolutions. 

The frames in \Fig{velocity_struct.fig} show the structure of the velocity flow around the null point for four of the experiments ($A$, $B$, and $C$ at t=7.5) and ($F$ at t=10.0) together with an isosurface of the strong current. The top frame represents experiment $A$, where it is seen that the velocity flow advects the plasma and magnetic field into the current sheet along a direction normal to the surface of the disk-shaped sheet, before expelling it from the narrow sides of the sheet close to the $y$-direction, in two high velocity reconnection jets. The figure shows that the current sheet is only slightly extended in the $z$-direction and therefore has the approximate shape of a tilted circular disk, but that the reconnection jets are primarily in the $y$-direction with only a very small velocity component in the $z$-direction. In other words for this set-up the structure of the velocity is similar to the rotationally symmetric null point cases previously investigated \citep{2007JGRA..11203103P,2007PhPl...14e2106P,2011A&A...529A..20G}.
The second frame represents experiment $B$. Here, both the current sheet and the reconnection jets are rotated relative to the imposed driver direction ($y$-direction). The velocity flow propagates more radially away from the null providing a more homogeneous flow structure. Ahead of the jet, the flows are seen to align with the imposed driving direction, showing how the initial slow perturbation is imposed on the system, but is then subsequently distorted as the current sheet aligns with the WE direction. A similar picture is seen for experiment $C$ (third frame), where it is also clear that there are differences in the alignment of the current sheet and the outflow velocity. The current is still aligned with the WE direction, while the tension forces are strongly influenced by the orientation of the imposed driving, leading to an asymmetry in the overall structure. Finally the bottom frame shows the same for experiment $F$, but for the later time, t=10. This is the time where the current reaches its peak value in a limited region around the null. But, as it is seen in this figure, the current accumulates inwards from the $z$-boundaries towards the null point as time progresses, rather than being peaked at the null as in A-C. The strong current we see for this event therefore does not represent a growing current sheet centred at the null. The reason for this is linked to the structure of the magnetic field. As the field is sheared in the WE direction, the perturbations rapidly propagate out along the SE direction reaching the $z$-boundary where the field line motion is frozen due to the imposed boundary conditions. This gives a characteristic shear motion across the fan plane building up strong currents close to the boundaries. For the velocity, a flow pattern is soon developed where the peak value is reached in four patches located around the null point lying close the $x=0$ plane. These are not associated with the jets driven by reconnection at the null, but are effects of enhanced diffusion away from the null due to the continuously imposed shearing and the accumulation of strong current close to the $z$-boundary. Note that the collapse around the null point is inhibited in this case because there is no component of the shear disturbance along the SE direction. Rather it is exactly aligned with the WE direction, and thus the Lorentz forces that trigger the collapse of the null are weak owing to the weak magnetic field in the shearing plane. 

\subsection{Reconnection rate}
\begin{figure}
{\hfill \includegraphics[width=0.5\textwidth]{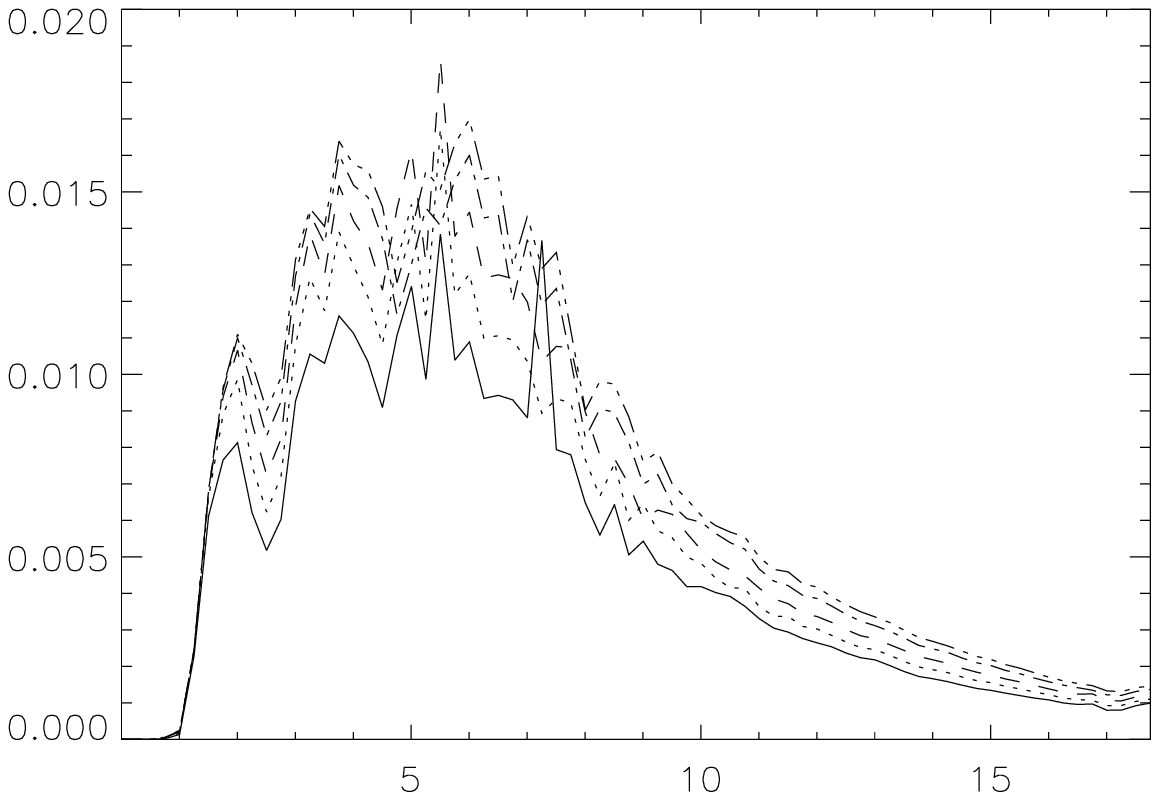} \hfill}

{\hfill \includegraphics[width=0.5\textwidth]{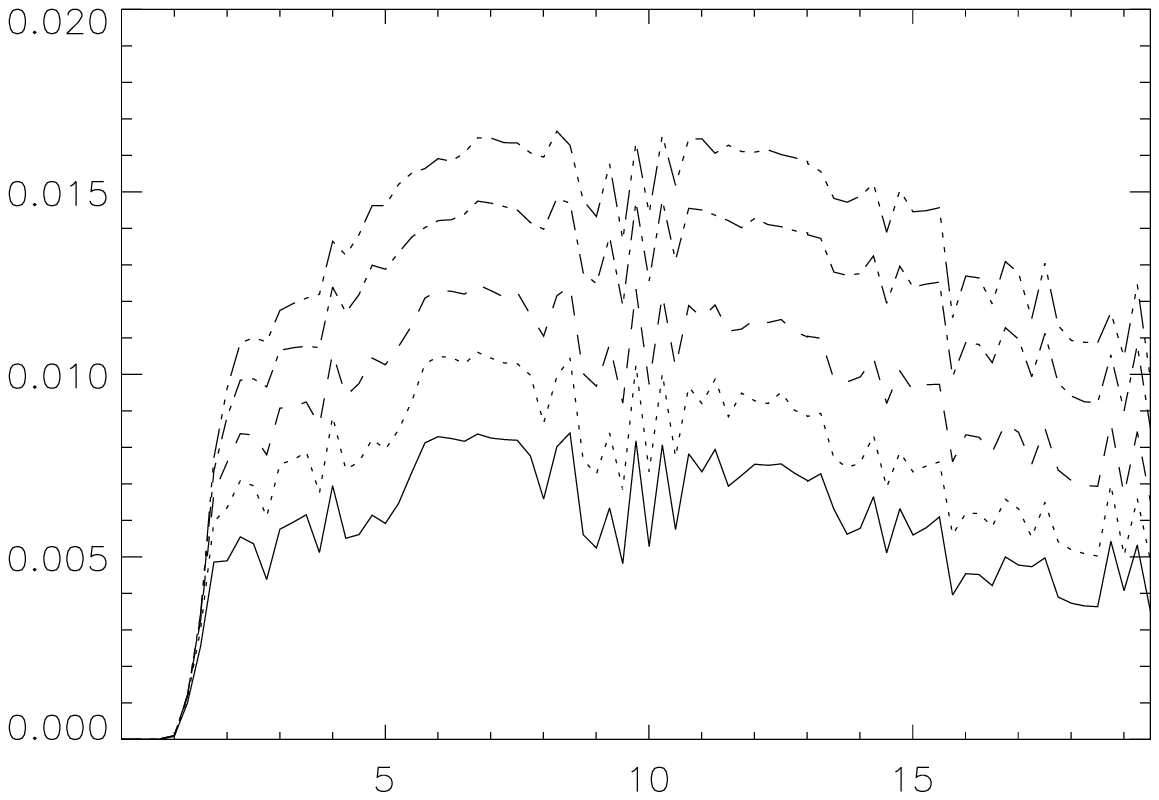} \hfill}

{\hfill \includegraphics[width=0.5\textwidth]{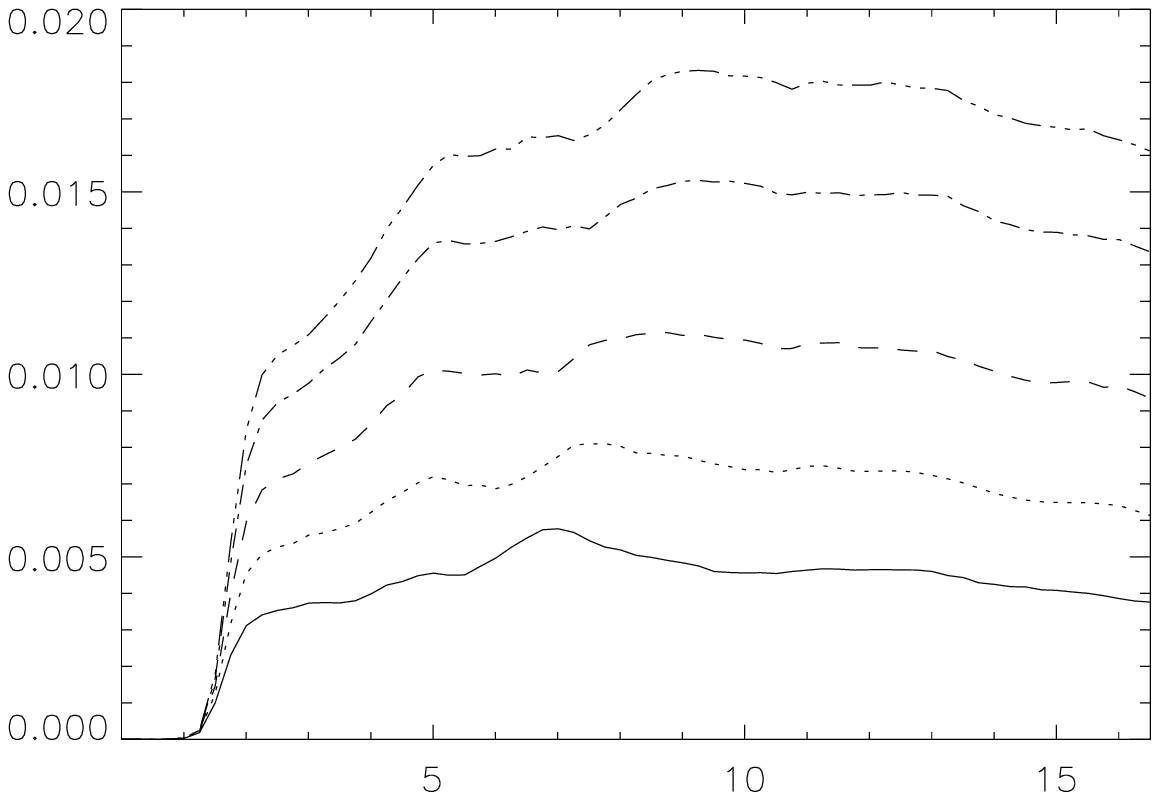} \hfill}
\caption[]{\label{epar_evolution_single.fig} The peak integrated parallel electric field for experiments A (top), C (middle) and F (bottom) for different values of the integration distance, $L$, as a function of time. The lines represents $L$: 0.5 is full line, 0.75 is dotted line, 1.0 is long dashed line, 1.25 is dot-dashed line and 1.4 is dashed triple dotted line.}
\end{figure}
An important parameter to measure is the reconnection rate. This gives a simple way to compare the efficiency of converting free magnetic energy into different types of energy for the different relative orientations. In three dimensions the reconnection rate is determined as the peak value of the integrated parallel electric field along all field lines passing through the diffusion region \citep{1988JGR....93.5547S}. Here this integral in general reaches its maximum value for a field line in the fan plane. To avoid numerical problems integrating field lines very close to the null we integrate 
\EQ
\label{int_E_par.eq}
\Psi =\int E_\| {\rm d}l = \int_{\epsilon}^{L} {\eta \JJ \cdot \BB \over |\BB|} ~{\rm d}l,
\EN
where $\epsilon$ is a small distance from the null point --- here we take a value of 0.01. $L$ is a maximum distance to integrate to avoid problems from the interaction with the imposed boundary conditions.

A simple way to check that consistent results are obtained from the experiments is to change the value of $L$. If the current is localised around the null point, then the integrated parallel electric field should converge towards a constant value as $L$ is increased. When significant currents are present far from the null, the integral will continue to increasing with larger $L$ values. To investigate this the integral has been calculated for 5 different values of $L$ (= 0.5, 0.75, 1.0, 1.25 and 1.4).
\Fig{epar_evolution_single.fig} shows the development of $\Psi_{max}$ for three of the experiments (A, C, F), where the different lines represent the different integration lengths $L$. We see that only for the A and B cases do we find a convergence of $\Psi_{max}$ for increasing $L$. For the other cases we found a continuously growing value of the parallel electric field. This shows that the continuous driving experiment is strongly limited in deriving reconnection rates as a function of $\theta$. This limitation is related to the linear structure of the magnetic field and the imposed boundary conditions. 

\section{Impulsive driving}\label{impulsive.sec}
In this section we consider the case where the boundary driving is imposed in an impulsive manner. That is, the boundary shearing is applied for some fixed period of time before being turned back to zero. We apply a driver with the spatial profile given in \Eq{driver.eq} but with the amplitude now being localised in time, via
\begin{equation}\label{incompprof}
V_d(t) = v_0 \left( \left(\frac{t-\tau}{\tau}\right) ^4 -1 \right) ^2 \qquad
\qquad 0 \leq t \leq 2\tau.
\end{equation}
We take $v_0=0.01$ and $\tau=1.5$ so that the driving switches off at $t=3$. All of the other parameters are the same as in the continuously driven simulations described above, except that in this case to resolve the current sheet we only require to use a grid resolution of $96\times 128^2$.

The evolution of the system is qualitatively similar to that described by \cite{2007PhPl...14e2106P} who investigated the rotationally symmetric case. As with the continuously driven case described above the impulsive shear driving results in a magnetic field disturbance that propagates into the domain, focussing around the null point. The difference here is that for all cases a current layer forms in this region around the null, with the current maximum located at the null point. The magnitude and dimensions of the current layer increase until the driving is switched off, after which the resistive diffusion gradually dissipates the current and the magnetic field relaxes back towards the orthogonal null point configuration of the initial field.  
For longer runs oscillations in both angle and current magnitude were observed, as the initial perturbation was bounced between the closed domain boundaries and thereby re-entered the area around the null point several times.

Here we have repeated the impulsive driving of the system with the same initial conditions as described in \Sec{setup.sec}. The simulation was run with $\theta=0,15,30,45,60,75,90$. Broadly speaking the results concerning the geometry of the current layer are similar to those found for the continually driven system. In particular, the extension of the current layer on large scales far away from the null is selected by the direction of the SE. However, locally the current flows through the null orthogonal to this, in the WE direction. 

\begin{figure}
\includegraphics[width=0.24\textwidth]{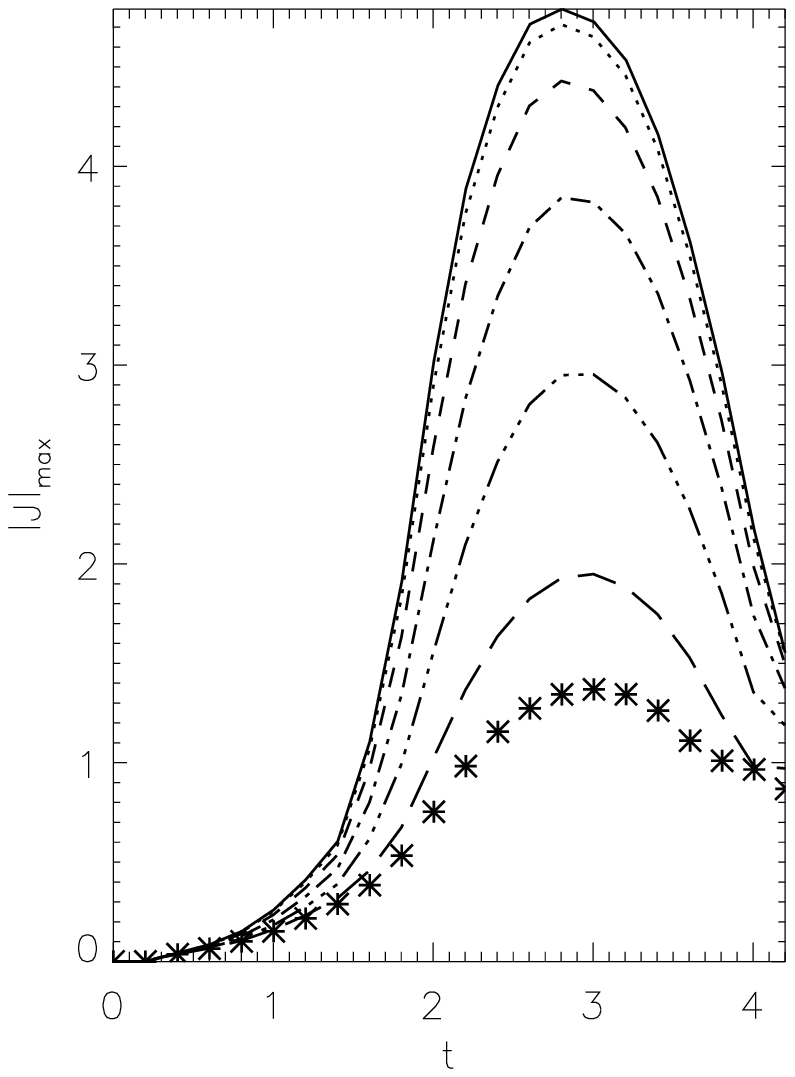}
\includegraphics[width=0.24\textwidth]{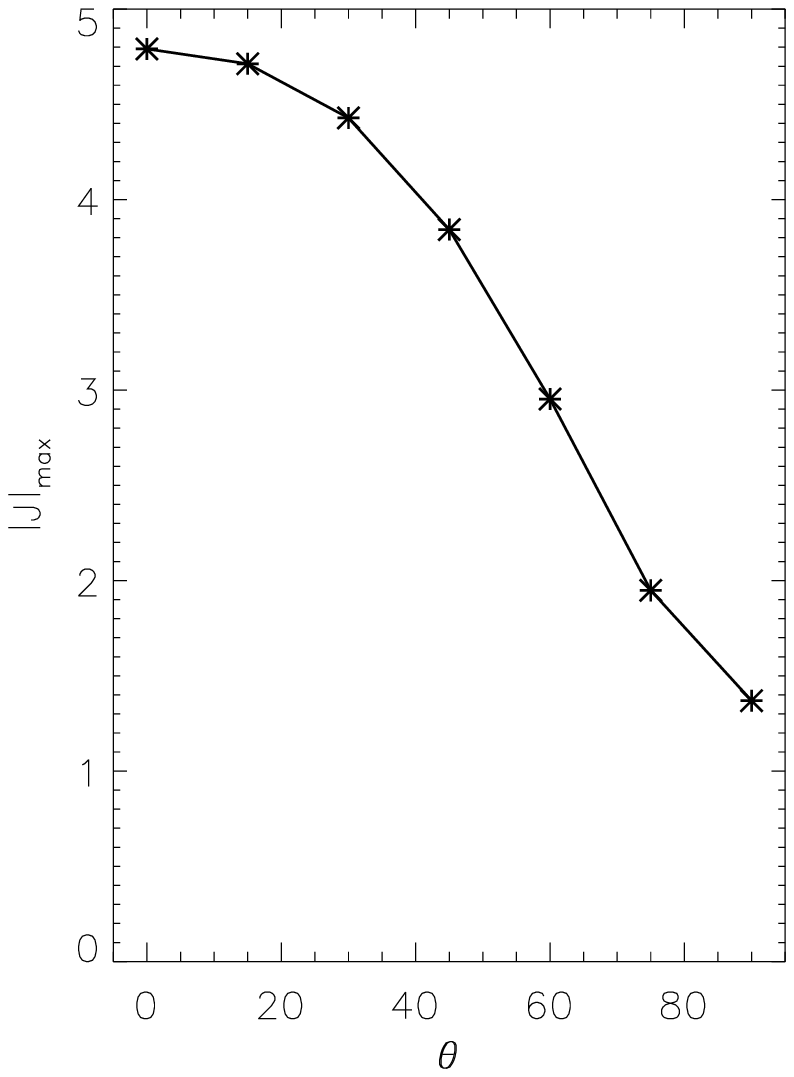}
\caption[]{\label{Jimpulsive.fig} Left: evolution of the spatial maximum of the current density with time for the different impulsive driving runs ($\theta=0$ is the solid line, $\theta=15$ dotted $\theta=30$ dashed, $\theta=45$ dot-dashed, $\theta=60$ triple-dot-dashed, $\theta=75$ long dashed, $\theta=90$ stars). Right: spatial and temporal maximum of $|{\bf J}|$ as a function of $\theta$.}
\end{figure}
\begin{figure}
\includegraphics[width=0.24\textwidth]{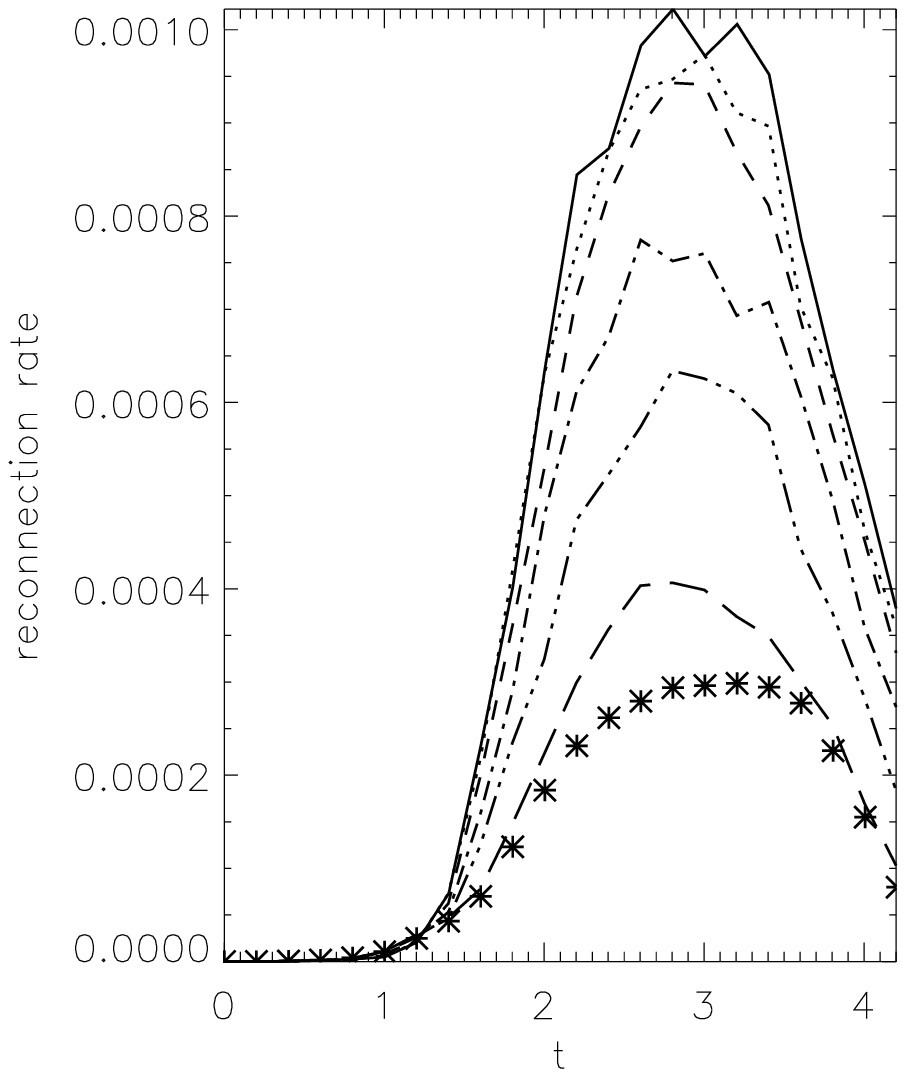}
\includegraphics[width=0.24\textwidth]{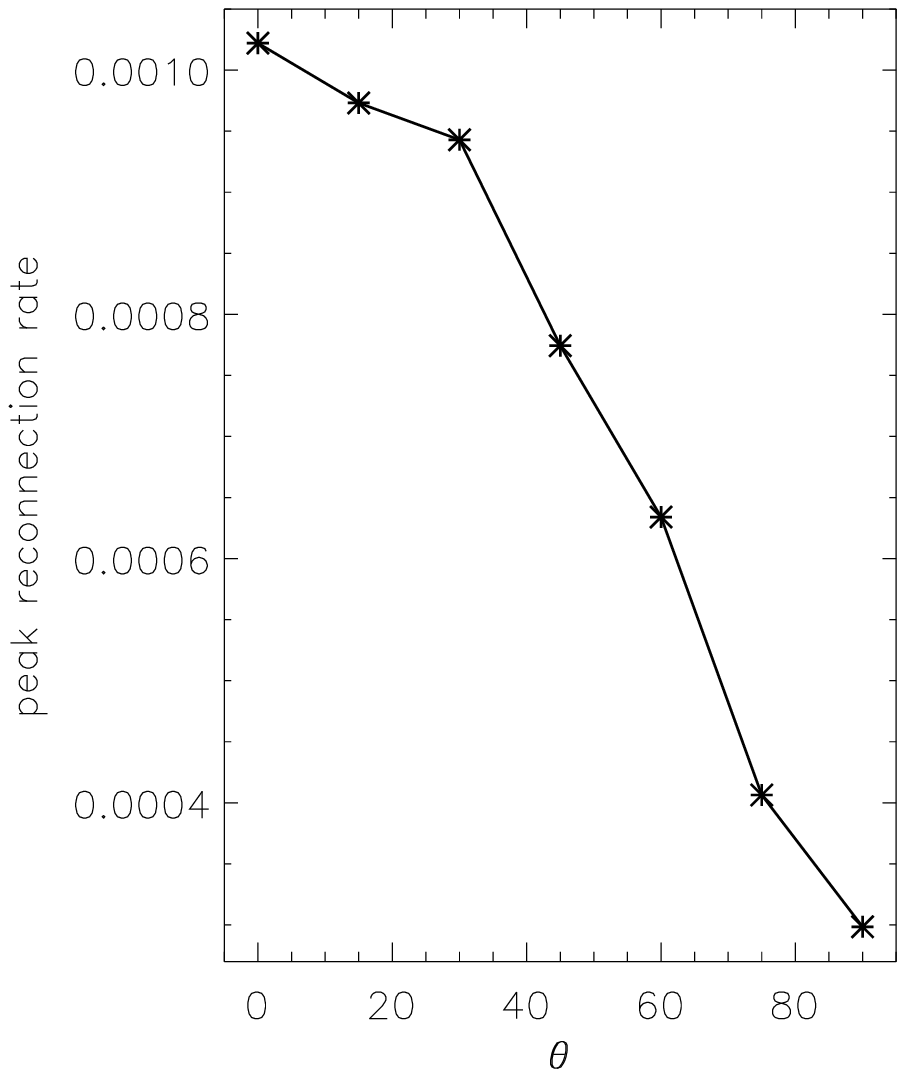}
\caption[]{\label{E_par_impulsive.fig} Left: evolution of the spatial maximum of the integrated parallel electric field (i.e.~the reconnection rate) with time for the different impulsive driving runs ($\theta=0$ is the solid line, $\theta=15$ dotted $\theta=30$ dashed, $\theta=45$ dot-dashed, $\theta=60$ triple-dot-dashed, $\theta=75$ long dashed, $\theta=90$ stars). Right: spatial and temporal maximum of the reconnection rate as a function of $\theta$.}
\end{figure}

The only simulation runs for which the above does not adequately describe the current sheet geometry are when $\theta=0^\circ$ and $\theta=90^\circ$. For these simulations the shear driving is exactly along the SE ($\theta=0^\circ$) or WE ($\theta=90^\circ$) directions. In each case the current vector at the null is perpendicular to the plane of the shear, with the current layer extending along this direction. The magnitude of the peak current density, however, is very different in these two simulations. Specifically, when $\theta=0^\circ$ and the driving is along the SE direction, the collapse is much more efficient due to the higher Lorentz forces, and the peak current is much higher. Indeed, there is a continuum of values for the peak current for the different runs, with the peak current decreasing as $\theta$ is increased, as shown in \Fig{Jimpulsive.fig}. The same pattern in seen in the outflow jet velocity (see \Tab{driver.tab}), with the peak jet velocity decreasing as $\theta$ increases and the current density decreases.

Turning now to the reconnection rate, we again observe a similar behaviour as for the peak current density. As shown in the left frame of \Fig{E_par_impulsive.fig}, the reconnection rate increases steadily as the null is sheared, reaching a maximum value around the time that the driving ceases ($t\approx 3$), before then decreasing. Examining the dependence on the angle $\theta$ of the maximum reconnection rate, we find the peak value of the reconnection rate over time decreases for increasing $\theta$ (right frame of \Fig{E_par_impulsive.fig}). 

\section{Discussion}
\label{discu.sec}

Previous investigations of rotationally symmetric null points found the current to accumulate in the direction perpendicular to the imposed stress of the spine axis. This is expected and is supported using simple arguments, based on Amp{\` e}re's law. When the rotational symmetry of the initial magnetic null point is broken, and the WE is perpendicular to the direction of the spine stress, the accumulating current sheet extends steadily further along the WE direction as this eigenvalue decreases in relative magnitude \citep{2010A&A...512A..84A}. In this paper we showed that the current continues to accumulate along the WE direction even for cases where the driver direction is no longer perpendicular to this plane. This can be understood as follows: It is the Lorentz force that is responsible for initiating the collapse of the magnetic field around the null to form the current layer (while the plasma pressure force acts to oppose the collapse). Thus, so long as some component of the shear disturbance acts in the SE direction, the magnetic field will preferentially collapse in this plane since the Lorentz force will be strongest in this plane. The collapse of the null in the SE plane naturally leads to a current vector perpendicular to that plane, i.e. along the WE direction. 
However, this collapse to form a current layer at the null does not always dominate the dynamics. When the angle between the SE and the driving direction reaches about $70^\circ$ there is a change in the behaviour of the current accumulation found for the continuous driving case. For these situations the imposed stress propagates out along the field lines away from the null in a direction that is near perpendicular to the imposed stress, and instead it accumulates close to the $z$-boundaries. This effect is enhanced by the imposed boundary conditions, that freeze the field line footpoints at the side boundaries preventing the stress from propagating away from the null region. As a result of the two competing effects, the current layer has a `zig-zag' shape when viewed along the $x$-direction looking down on the fan plane, see \Fig{current_structure.fig}. This shows that having an asymmetric 3D null and imposing a shear perturbation, a strong current will not necessary accumulate at the null point, depending on the orientation of the shearing with respect to the null structure. 

Comparing the continuous and impulsive driving experiments, we see that it is only when the system is driven continuously that the current begins to accumulate away from the null. In the impulsive driving case the current is always focussed at the null, as there is no time for the information to communicate with the domain boundaries before the pulse has passed the null point region. One therefore does not expect to see this domain effect for this type of driving. This is also seen in the fact that the peak current reached in the two experiments differs with a factor of 6, with the systematic driving being able to impose a much stronger deformation of the magnetic field around the null.

Magnetic reconnection is often associated with fast plasma outflows. Here we find that knowing the structure of the magnetic field in the vicinity of the null point provides insight as to which directions one can expect to find the reconnection jets. The current aligns locally along the WE direction, with the reconnection jet being ejected mainly perpendicular to this along the SE direction. As the angle between the SE and the driver reaches near perpendicular, the null point is stressed in a way where no jet is formed due to a current concentration around the null, but the reconnection speed is still very high. Instead four jet-like structures are located in the old fan plane and are due to the influence from the imposed boundary conditions. This is important in that it is not obvious that signatures of fast magnetic reconnection in relation with null reconnection will always show up as narrow single jet regions around a common diffusion region. 

The evolution of the reconnection rate is not so `smooth' as for ${\bf J}$ (see Figs.~\ref{epar_evolution_single.fig}, \ref{E_par_impulsive.fig}) due to the difficulty in obtaining this quantity. This is because $E_\|$ is typically highest in the weak field regions (see the above discussion regarding ${\bf J}$), so that obtaining a high field line density in this region to accurately capture the spatial maximum is not straightforward. The results obtained show a clear difference between the impulsive and sustained driving. For the impulsive driving, we find that the reconnection rate decreases as we increase $\theta$ mirroring the behaviour of the peak current (which is natural since $E_\|=\eta J_\|$). In this case the effect of the boundaries cannot influence the dynamics at the null region before the peak of the reconnection process has occurred. On the other hand, for the sustained driving only the cases with small $\theta$ values have the current concentrated around the null such that actual reconnection rates for the null reconnection can be obtained. For the remaining cases we convincingly showed that the contribution to determination of the reconnection rate continues to the edge of the domain. To first order this is an effect of the imposed boundary conditions and the linear structure of the magnetic field used. It will therefore be interesting and important to redo this investigation with a different magnetic configuration that limits the influence of the faraway boundary conditions on the local evolution at the null. By contrast to the continuously driven case, the time-limited driving gives a current pulse that always becomes localised around the null. The reconnection rate therefore does not depend on the integration length $L$ along the field line (so long as $L$ is taken to be sufficiently large) as was seen for the continuous driving case.

\section{Conclusion}
\label{conc.sec}

Recent investigations of magnetic reconnection at single 3D null points clearly show the importance of the combination between the magnetic field structure and the imposed driving. While the collapse of the null -- and therefore the formation of the current layer -- depends strongly on the structure of the magnetic field, the driving seems only to be the provider of stress necessary to foster the collapse and the subsequent reconnection process. Both the current density in the sheet and the associated reconnection jet velocity are strongly related to the relative orientation between the magnetic field structure and the driver, with the cases where the null point's SE direction align with the stress of the spine axis providing the largest amplitudes of these directly measurable physical variables.

In order to model coronal magnetic fields, one can use different techniques to extrapolate the photospheric field into the corona, employing potential or non-linear force free extrapolations. Depending on which approach is used the local structure of the magnetic field around the null will change. Therefore it is difficult to give a precise description of the  dynamical evolution of the magnetic field that would be expected when it is exposed to external (boundary driven) stresses. However, combining observations for the flow direction (when possible) and photospheric stressing patterns with field extrapolations, one may be able to predict the local behaviour of the 3D null point using the results discussed in the recent papers on 3D null reconnection \citep{2007JGRA..11203103P, 2007PhPl...14e2106P,2010A&A...512A..84A, 2011A&A...529A..20G}. This is what was done by \citet{2009ApJ...700..559M}, but for a case with an indirect stressing of the magnetic null that led to the collapse that facilitated the ongoing magnetic reconnection. A similar indirect perturbation has not been investigated for generic null points, and therefore direct predictions of the null evolution are uncertain.

To provide a better working hypothesis for null point behaviour more experiments need to be conducted, using magnetic field configurations that are not linearly dependent on space coordinates to better see how non-linear structural changes of the magnetic field will influence the reconnection process. It is vital to repeat the sustained driving case studied here with a different initial null point configuration that limits the implications of the boundaries and allows for an independent investigation of the reconnection rate at the null point. Furthermore, in order facilitate more direct comparison of the results with realistic situations we need to make scaling predictions on the behaviour of the null evolution. This is a very difficult task as it requires that we repeat experiments with the magnetic resistivity changing over several orders of magnitude -- a process requiring a substantially higher numerical resolution than the present experiments \citep{2011A&A...529A..20G}.

\bigskip
{\it Acknowledgements:}
Computing time was given by the DCSC at university of Copenhagen. Support by the European Commission through the Solaire Network (MTRNCT-2006-035484) is gratefully acknowledged. DP acknowledges financial support from the Royal Society.\\

\bibliographystyle{aa} 
\bibliography{ref} 
\end{document}